\documentclass[aps,prd,onecolumn,nofootinbib]{revtex4-2}
\usepackage{amsmath,amssymb}

\begin{document}

	\title{Klein--Gordon Dynamics from Intrinsic Phase Periodicity}

	\author{Emiliano Puddu}
	\email[]{epuddu@liuc.it}
	\affiliation{Università Carlo Cattaneo -- LIUC, 21053 Castellanza, Italy}

	\date{\today}
	
	\begin{abstract}
		This work develops a phase-based formulation of relativistic wave dynamics, demonstrating that the Klein--Gordon equation emerges naturally from the foundational assumption of intrinsic phase periodicity in material fields. Mapping the phase directly onto the classical action, we postulate that localized excitations possess an invariant rest-frame oscillation governed by a proper frequency $\omega_0$. This physical condition establishes an operational mass-frequency relation, $m = \hbar \omega_0 / c^2$, without requiring rest energy as an independent, axiomatic input. We show that the Klein--Gordon equation arises as the minimal local, linear, Lorentz-invariant field equation compatible with this internal phase structure. Within this framework, mass acts as an intrinsic frequency scale governing wave propagation, and relativistic kinematics is fully recovered as a structural consequence of phase coherence. This approach provides a unified wave-mechanical interpretation where particle dynamics maps onto the group velocity of dispersive wave packets, offering an intuitive account of free propagation, dispersion, and tunneling across potential barriers.
	\end{abstract}
	
\maketitle

\section{Introduction}

Relativistic quantum mechanics conventionally introduces the Klein--Gordon equation as a formal quadratic generalization of the Schrödinger equation, obtained by applying the standard quantum mechanical substitution rules to the relativistic energy-momentum relation~\cite{greiner1990,peskin1995}. While this textbook derivation ensures compatibility with special relativity, it treats the algebraic structure of the field equation and the geometric properties of spacetime as primary, independent axioms. The wave-mechanical properties of the field---most notably its phase structure---are subsequently recovered as mathematical consequences of the formal machinery.

Historically, however, the connection between relativistic kinematics and wave properties was discovered through a different logical sequence. In his foundational work, de Broglie postulated that any localized material particle possesses an internal periodic phenomenon, setting up an intrinsic clock whose frequency is tied to the particle's mass~\cite{debroglie1924}. This insight paved the way for wave mechanics, showing that relativistic kinematics and wave propagation are intimately linked at a structural level. 

Building on this historical perspective, recent investigations have explored the possibility of reinterpreting relativistic kinematics not as an independent geometric background, but as an emergent manifestation of phase coherence and intrinsic periodicity in localized wave states~\cite{muller2014, puddu2026kinematics}. By treating phase as the primary invariant quantity, it has been demonstrated that proper time operationally tracks the cycle accumulation of an internal clock, while the Minkowski metric arises naturally as the unique quadratic structure required to sustain phase coherence across moving frames.

The present work extends this phase-based approach from pure kinematics to relativistic wave dynamics. Rather than taking the Klein--Gordon equation as a primary postulate or a byproduct of a geometric metric, we demonstrate that it can be systematically derived from the requirement of intrinsic phase periodicity. Under the standard identification of the phase $\Phi$ with the classical action $S$ via $\Phi = S/\hbar$, we assume that a localized physical state admits an invariant rest-frame oscillation characterized by a proper frequency $\omega_0$. This fundamental condition yields the mass-frequency relation $m = \hbar \omega_0 / c^2$ as a direct, structural identity, eliminating the need to introduce rest energy as an independent phenomenological parameter.

We show that the Klein--Gordon equation emerges as the minimal local, linear, and Lorentz-invariant field equation compatible with this intrinsic periodicity. Mass is thus reinterpreted as an internal frequency scale that governs the dispersive propagation of the field. This formulation does not alter the empirical predictions of relativistic quantum mechanics, but it provides a self-consistent physical interpretation where particle behavior is entirely subordinated to wave dynamics. Kinematic properties emerge from phase evolution, and particle trajectories map onto the group velocity of dispersive wave packets.

This perspective offers a unified interpretation of several relativistic phenomena. Free wave propagation and dispersion follow directly from the underlying phase structure, while the transition between oscillatory and evanescent behavior provides an intuitive wave-mechanical account of tunneling across potential barriers, viewed here as propagation below the intrinsic cutoff frequency. Furthermore, the nonrelativistic limit follows naturally: removing the fast intrinsic rest-phase straightforwardly recovers the Schrödinger equation for the slowly varying envelope, highlighting how nonrelativistic quantum mechanics inherits its structure from a deeply relativistic phase architecture.

The paper is organized as follows. Section~II reviews the physical foundations of intrinsic phase periodicity and its relation to mass. Section~III details the systematic derivation of the Klein--Gordon equation from these phase coherence conditions. Section~IV analyzes free wave propagation and the group velocity of wave packets. Section~V explores applications to physical potentials, focusing on dispersion and tunneling phenomena. Finally, Section~VI evaluates the implications and limitations of this framework, followed by concluding remarks in Section~VII.

\section{Intrinsic Phase Periodicity and Mass--Frequency Relation}

\subsection{Phase as action}

Rather than altering the empirical predictions of relativistic quantum mechanics, this work aims to provide a unified, phase-based formulation where the Klein–Gordon equation flows from a single structural principle. To this end, let us consider a complex scalar field $\psi(x)$ representing material excitations. Within a wave-mechanical framework, the phase $\Phi(x)$ of the field acts as a primary physical quantity, mapping directly onto the classical action $S(x)$ through the fundamental relation:
\begin{equation}
	\Phi(x) = \frac{S(x)}{\hbar}.
\end{equation}
This identification traces back to the Hamilton–Jacobi formulation of mechanics and stands at the very core of wave mechanics, as originally envisioned in the foundational insights on matter waves  \cite{debroglie1924}.

The spacetime gradient of the phase explicitly determines the local energy-momentum distribution of the field:
\begin{equation}
	\partial_\mu \Phi = \frac{1}{\hbar} \partial_\mu S = \frac{1}{\hbar} p_\mu,
\end{equation}
where $p_\mu$ denotes the energy-momentum four-vector. Governed by this relation, the phase establishes an immediate, structural bridge between geometric wave properties and classical dynamical quantities.

\subsection{Proper-time periodicity and invariant frequency}

Following de Broglie's historical hypothesis that massive entities possess an underlying periodic nature, we assume that any localized material state carries an intrinsic rest-frame phase oscillation governed by a proper angular frequency $\omega_0$. In this co-moving frame, the phase evolves harmonically as:
\begin{equation}
	\Phi_0=-\omega_0 t_0 .
\end{equation}
Since the rest-frame coordinate time maps identically onto the proper time along the system's worldline, this evolution generalizes into the manifest Lorentz scalar relation:
\begin{equation}
	d\Phi=-\omega_0 d\tau .
\end{equation}
Rather than representing an independent dynamic postulate, this condition serves as the strict covariant formalization of de Broglie's internal-clock concept \cite{debroglie1924}. Operationally, Lorentz invariance dictates that such an internal periodicity must be fundamentally parameterized by the proper time, establishing that the total phase accumulated along a physical trajectory directly counts the cycles of this localized clock.

Under the standard mapping $\Phi = S/\hbar$, this phase accumulation rule reflects a specific constraint on the classical action differential:
\begin{equation}
	dS = -\hbar \omega_0\, d\tau.
\end{equation}

\subsection{Emergence of the mass scale}

In relativistic mechanics, the infinitesimal action for a free particle is conventionally written as:
\begin{equation}
	dS = -mc^2\, d\tau.
\end{equation}
Matching this canonical definition against the constraint imposed by the intrinsic phase periodicity straightforwardly yields the identity:
\begin{equation}
	\hbar \omega_0 = mc^2.
\end{equation}

This identification structurally captures the particle mass as the intrinsic rest-frame frequency of the wave field. Within this perspective, mass ceases to be an abstract coupling constant or an isolated phenomenological parameter; it emerges instead as a direct operational measure of the internal phase oscillation.

Consequently, the invariant frequency $\omega_0$ sets the foundational scale governing field propagation. As developed in the subsequent sections, $\omega_0$ acts as a physical cutoff threshold in the relativistic dispersion relation, below which traveling, oscillatory solutions become evanescent. This mapping provides a transparent wave-mechanical interpretation of mass, linking it directly to the behavior of waves propagating through dispersive media.

Furthermore, under the identification of phase with action, the classical relativistic Hamilton--Jacobi equation,
\begin{equation}
	\partial_\mu S \, \partial^\mu S = m^2 c^2,
\end{equation}
becomes deeply encoded within the geometric gradient of the phase field. This non-linear differential constraint provides the necessary conceptual and structural bridge required to derive the corresponding linear covariant field equation.

\section{Minimal Covariant Dynamics from Intrinsic Periodicity}

\subsection{Guiding principles}

The identity $\hbar \omega_0 = mc^2$ established in the previous section explicitly maps the mass scale of a material excitation onto an intrinsic rest-frame frequency. A crucial question naturally follows: does this foundational phase architecture uniquely determine the form of the corresponding relativistic field equation?

Our objective is not to postulate the Klein--Gordon structure from the outset as an independent axiom, but rather to demonstrate that it emerges as the minimal local, linear, and Lorentz-invariant scalar equation compatible with this phase-centric framework. To formalize this approach, we require the dynamics of the scalar field $\psi(x)$ to satisfy four guiding constraints:

\begin{enumerate}
	\item \emph{Locality}: the differential equation must exclusively relate the field and its spacetime derivatives evaluated at the same event;
	\item \emph{Linearity}: the principle of superposition must hold rigorously in the free-field regime;
	\item \emph{Lorentz invariance}: the structural form of the equation must remain invariant across all inertial frames;
	\item \emph{Compatibility with intrinsic phase periodicity}: the framework must admit free wave solutions whose phase accumulates according to the invariant scalar relation
	\begin{equation}
		d\Phi = -\omega_0\, d\tau,
	\end{equation}
	where $\omega_0$ represents the invariant rest-frame frequency scale.
\end{enumerate}

These physical requirements heavily restrict the available dynamical configurations. For a scalar field, Lorentz invariance demands that spacetime derivatives enter through scalar contractions constructed from the gradient operator $\partial_\mu$. The simplest non-trivial operator satisfying this condition is the d'Alembertian,
\begin{equation}
	\Box \equiv \partial_\mu \partial^\mu
	= \frac{1}{c^2}\frac{\partial^2}{\partial t^2} - \nabla^2,
\end{equation}
evaluated under the standard $(+,-,-,-)$ metric signature.

We are therefore led to seek the minimal linear local field equation of the general form:
\begin{equation}
	\left(\Box + \lambda^2\right)\psi = 0,
	\label{eq:minimal_scalar_eq}
\end{equation}
where $\lambda$ represents a constant with dimensions of inverse length. The analytical task now reduces to determining the physical value of $\lambda$ strictly from the field's underlying phase periodicity.

\subsection{Plane-wave solutions and dispersion relation}

To illuminate the physical content encoded within Eq.~(\ref{eq:minimal_scalar_eq}), let us consider free plane-wave configurations defined by the standard ansatz:
\begin{equation}
	\psi(x,t) = A e^{i(\mathbf{k}\cdot\mathbf{x}-\omega t)},
	\label{eq:plane_wave_ansatz}
\end{equation}
where $A$ denotes a constant amplitude, $\mathbf{k}$ the wavevector, and $\omega$ the angular frequency.

Direct substitution of Eq.~(\ref{eq:plane_wave_ansatz}) into the minimal scalar equation~(\ref{eq:minimal_scalar_eq}) yields:
\begin{equation}
	\left(-\frac{\omega^2}{c^2} + k^2 + \lambda^2\right)\psi = 0,
\end{equation}
which straightforwardly determines the corresponding dispersion relation:
\begin{equation}
	\omega^2 = c^2 k^2 + c^2 \lambda^2.
	\label{eq:dispersion_lambda}
\end{equation}

This equation matches the exact algebraic form of a relativistic dispersion law governed by a non-vanishing frequency offset. Crucially, in the limit of a spatially uniform excitation, the wavevector vanishes, giving:
\begin{equation}
	k=0 \qquad \Rightarrow \qquad \omega = c\lambda.
\end{equation}
Because the $k=0$ configuration corresponds precisely to the rest frame of the localized wave excitation, consistency with the intrinsic phase periodicity developed in Sec.~II demands that this frequency match the proper frequency scale:
\begin{equation}
	c\lambda = \omega_0.
\end{equation}
It follows immediately that $\lambda = \omega_0/c$, allowing Eq.~(\ref{eq:minimal_scalar_eq}) to be rewritten as:
\begin{equation}
	\left(\Box + \frac{\omega_0^2}{c^2}\right)\psi = 0.
	\label{eq:kg_omega0}
\end{equation}

Incorporating the operational mass--frequency relation $\hbar \omega_0 = mc^2$, this field equation assumes its standard textbook representation:
\begin{equation}
	\left(\Box + \frac{m^2 c^2}{\hbar^2}\right)\psi = 0,
	\label{eq:kg_standard}
\end{equation}
which is exactly the Klein--Gordon equation \cite{klein1926,gordon1926,kragh1984}.

The vital insight here lies in the logical inversion of its interpretation: the parameter conventionally designated as a mechanical mass term emerges naturally as the square of an intrinsic, rest-frame oscillation frequency. Consequently, Eq.~(\ref{eq:kg_standard}) is not introduced by applying formal quantization rules to a classical particle trajectory, but is derived as the minimal covariant wave architecture compatible with foundational phase coherence.

\subsection{Rest-frame interpretation}

The physical clarity of Eq.~(\ref{eq:kg_omega0}) becomes especially prominent when evaluated within the co-moving rest frame of the excitation. Setting $\mathbf{k}=0$, the plane-wave ansatz~(\ref{eq:plane_wave_ansatz}) collapses to:
\begin{equation}
	\psi(\tau) = A e^{-i\omega_0 \tau},
\end{equation}
where the laboratory coordinate time coincides with the proper time $\tau$. Under these conditions, the covariant field equation~(\ref{eq:kg_omega0}) reduces identically to:
\begin{equation}
	\frac{d^2 \psi}{d\tau^2} + \omega_0^2 \psi = 0,
	\label{eq:proper_time_oscillator}
\end{equation}
which is the structural equation of a simple harmonic oscillator governed by the invariant angular frequency $\omega_0$.

This result carries deep conceptual weight. It demonstrates that the covariant field dynamics, when stripped of spatial propagation, is locally equivalent to the harmonic evolution of an intrinsic clock. Conversely, the Klein--Gordon equation~(\ref{eq:kg_omega0}) can be entirely understood as the unique Lorentz-covariant extension of this fundamental rest-frame oscillator equation to arbitrary inertial frames. Mass is thus fundamentally linked to a non-vanishing proper-time frequency, rather than being introduced as an external, non-wave parameter.

\subsection{Eikonal limit and Hamilton--Jacobi structure}

To verify the structural self-consistency of this formulation, let us examine the field in its eikonal representation:
\begin{equation}
	\psi(x) = A(x)e^{iS(x)/\hbar},
	\label{eq:eikonal_ansatz}
\end{equation}
where $A(x)$ acts as a slowly varying spatial amplitude and $S(x)$ represents the phase function identified with the classical action.

Substituting this ansatz into the Klein--Gordon equation~(\ref{eq:kg_standard}) and expanding the resulting differential terms in powers of $\hbar$, the leading-order contribution straightforwardly yields:
\begin{equation}
	\partial_\mu S\, \partial^\mu S = m^2 c^2.
	\label{eq:HJ_rel}
\end{equation}
This expression is precisely the relativistic Hamilton--Jacobi equation for a free particle.

Accordingly, in the short-wavelength limit, the field dynamics dictated by intrinsic phase periodicity recovers standard classical relativistic kinematics. The wave phase $S/\hbar$ determines the local wavevector and frequency modulation of the field, while its leading-order evolution is governed by the exact same invariant mass scale that controls proper-time phase accumulation. This demonstrates that the Klein--Gordon structure and the Hamilton--Jacobi relation are not independent theoretical ingredients; both reflect a single, underlying physical reality: the existence of an invariant phase periodicity built into material excitations.

\subsection{Interpretation of the mass term as a cutoff frequency}

The dispersion relation stemming from Eq.~(\ref{eq:kg_omega0}),
\begin{equation}
	\omega^2 = c^2k^2 + \omega_0^2,
	\label{eq:dispersion_final}
\end{equation}
admits an intuitive, purely wave-mechanical interpretation. The proper frequency $\omega_0$ functions as a strict lower bound for the system: propagating, traveling-wave solutions cannot exist in the regime where $\omega < \omega_0$. In this sense, the material field behaves exactly like a dispersive medium endowed with a sharp cutoff frequency.

This parallel provides a transparent physical picture of the mass term. Rather than representing an abstract, externally assigned particle property, mass acts as the intrinsic frequency scale built directly into the field's propagation laws. The presence of mass guarantees that even modes with vanishing spatial wavevector preserve a non-trivial temporal oscillation. Conversely, the massless limit corresponds to the erasure of this proper-time oscillation scale, straightforwardly recovering a purely lightlike dispersion relation:
\begin{equation}
	\omega = ck.
\end{equation}

Relativistic wave dynamics can therefore be viewed as the propagation of field excitations through a medium characterized by an intrinsic temporal scale. This perspective will serve as our foundation in the subsequent sections, where we explore free propagation, wave packet envelope dynamics, and behavior within external potentials in greater detail.

\section{Free Wave Propagation}

\subsection{Plane-wave solutions and dispersion relation}

Let us investigate the propagation characteristics governed by the Klein--Gordon equation derived in the previous section:
\begin{equation}
	\left(\Box + \frac{\omega_0^2}{c^2}\right)\psi = 0,
	\label{eq:kg_section4}
\end{equation}
where $\omega_0$ represents the invariant rest-frame frequency characterizing the field.

To analyze the fundamental solutions of this wave equation, we begin by examining plane-wave configurations of the form:
\begin{equation}
	\psi(x,t) = A e^{i(\mathbf{k}\cdot \mathbf{x} - \omega t)},
	\label{eq:plane_wave}
\end{equation}
under a constant complex amplitude $A$.

Applying the appropriate spacetime derivatives to the plane-wave ansatz~(\ref{eq:plane_wave}) yields the structural relations:
\begin{align}
	\frac{\partial^2 \psi}{\partial t^2} &= -\omega^2 \psi, \\
	\nabla^2 \psi &= -k^2 \psi,
\end{align}
which allow the d'Alembert operator to be written as a purely algebraic contraction:
\begin{equation}
	\Box \psi = \left(-\frac{\omega^2}{c^2} + k^2\right)\psi.
\end{equation}

Inserting this result back into the covariant field equation~(\ref{eq:kg_section4}) enforces the constraint:
\begin{equation}
	\left(-\frac{\omega^2}{c^2} + k^2 + \frac{\omega_0^2}{c^2}\right)\psi = 0,
\end{equation}
which establishes the underlying relativistic dispersion relation:
\begin{equation}
	\omega^2 = c^2 k^2 + \omega_0^2.
	\label{eq:dispersion}
\end{equation}

This quadratic relation reveals that even in the limit of spatial uniformity ($k=0$), the field sustains a non-vanishing temporal oscillation at the proper frequency $\omega_0$. This fundamental floor directly maps onto the intrinsic proper-time periodicity established as our core kinematic postulate.

\subsection{Group velocity and particle interpretation}

The dispersion relation~(\ref{eq:dispersion}) deeply shapes the propagation dynamics of localized field states. The group velocity $v_g$, which governs the transport of energy and modulation envelopes, is defined by:
\begin{equation}
	v_g = \frac{d\omega}{dk}.
\end{equation}

Expressing the dispersion law explicitly as a function of the wavevector, $\omega(k) = \sqrt{c^2 k^2 + \omega_0^2}$, differentiation with respect to $k$ straightforwardly yields:
\begin{equation}
	\frac{d\omega}{dk} = \frac{c^2 k}{\sqrt{c^2 k^2 + \omega_0^2}},
\end{equation}
which relates the group velocity directly to the total frequency content:
\begin{equation}
	v_g = \frac{c^2 k}{\omega}.
	\label{eq:group_velocity}
\end{equation}

By incorporating the canonical de Broglie identities $E = \hbar \omega$ and $p = \hbar k$, Eq.~({\ref{eq:group_velocity}}) transforms into:
\begin{equation}
	v_g = \frac{pc^2}{E}.
\end{equation}
Crucially, this expression matches the standard relativistic velocity of a classical material particle.

This exact mathematical equivalence supports a purely wave-mechanical interpretation of matter: localized physical excitations are wave packets whose collective translational motion is dictated by their underlying group velocity. Within this conceptual architecture, particle kinematics ceases to be an independent, axiomatic postulate; it emerges as a structural consequence of dispersive field propagation.

\subsection{Wave packets and dispersion}

Strictly speaking, a single monochromatic plane wave represents an idealized state distributed across all spacetime; physical localization requires a coherent superposition of multiple modes. Plane-wave solutions should therefore be understood as providing the local, point-like dispersion properties that dictate the evolution of physically real wave packets.

To construct a realistic localized state, we define a continuous superposition of plane waves:
\begin{equation}
	\psi(x,t) = \int a(k)\, e^{i(kx - \omega(k)t)} dk,
	\label{eq:wave_packet}
\end{equation}
where the spectral amplitude profile $a(k)$ is assumed to be sharply peaked around a dominant central wavevector $k_0$.

Performing a Taylor expansion of the frequency function $\omega(k)$ around this central value $k_0$ gives:
\begin{equation}
	\omega(k) \simeq \omega_0^{(k)} + (k-k_0)\left.\frac{d\omega}{dk}\right|_{k_0}
	+ \frac{1}{2}(k-k_0)^2 \left.\frac{d^2\omega}{dk^2}\right|_{k_0} + \cdots,
\end{equation}
where $\omega_0^{(k)} = \omega(k_0)$. While the first-order linear coefficient dictates the translation of the overall envelope at the group velocity $v_g$, the second-order quadratic term governs its structural deformation.

Differentiating the group velocity relation~(\ref{eq:group_velocity}) a second time with respect to $k$ yields the dispersion parameter:
\begin{equation}
	\frac{d^2 \omega}{dk^2} = \frac{c^2 \omega_0^2}{\left(c^2 k^2 + \omega_0^2\right)^{3/2}}.
\end{equation}

Because this second derivative remains strictly non-zero for any finite proper frequency ($\omega_0 \neq 0$), the wave packet inevitably undergoes continuous broadening over time. Physically, this shows that mass---acting through the identity $\hbar \omega_0 = mc^2$---directly controls the dispersion rate of the wave packet. Massive states correspond to fields endowed with a rich temporal architecture, meaning their localized profiles naturally spread during free propagation.

\subsection{Rest-frame limit and intrinsic oscillation}

The co-moving rest frame of the field excitation is operationally recovered in the long-wavelength limit where $k \to 0$. Under this condition, the relativistic dispersion relation~(\ref{eq:dispersion}) simplifies to:
\begin{equation}
	\omega = \omega_0,
\end{equation}
and the spatial dependence of the plane-wave solution drops out entirely, leaving:
\begin{equation}
	\psi(\tau) = A e^{-i\omega_0 \tau}.
\end{equation}

This reduction recovers the foundational rest-frame phase clock introduced in Sec.~II. It highlights that the free-field dynamics governed by the Klein--Gordon structure is entirely consistent with the requirement that each material excitation carries an invariant internal clock rate.

\subsection{Interpretation as propagation in a dispersive medium}

From a physical standpoint, the mathematical form of the dispersion relation~(\ref{eq:dispersion}) is identical to that of waves propagating through a medium characterized by a sharp cutoff frequency. Within this analogy, the proper frequency $\omega_0$ functions as a strict lower energy threshold: any mode operating at a frequency $\omega < \omega_0$ is forbidden from traveling as an oscillatory wave solution.

This parallel provides a transparent physical reading of the mass term. Rather than representing an abstract, externally assigned mechanical parameter, mass manifests as the intrinsic frequency scale built directly into the field's propagation laws. The presence of a non-vanishing mass guarantees that even modes stripped of spatial momentum preserve an active temporal oscillation. Conversely, the massless limit ($\omega_0 \to 0$) erases this proper-time scale, straightforwardly recovering a lightlike, non-dispersive dispersion relation:
\begin{equation}
	\omega = ck.
\end{equation}

Relativistic wave mechanics can therefore be understood as the study of field excitations moving through a medium endowed with an intrinsic temporal scale. This perspective serves as our foundational baseline for the following sections, where we explore envelope transformations, current densities, and interactions with external potentials.

\section{Nonrelativistic Limit}

\subsection{Separation of the intrinsic phase}

Let us now investigate the nonrelativistic behavior of the Klein--Gordon dynamics established in the preceding sections:
\begin{equation}
	\left(\Box + \frac{\omega_0^2}{c^2}\right)\psi = 0,
	\label{eq:kg_nr_start}
\end{equation}
where $\omega_0$ represents the intrinsic rest-frame frequency threshold of the field.

A prominent feature governing the solutions of Eq.~(\ref{eq:kg_nr_start}) is the persistence of rapid, baseline oscillations at the frequency $\omega_0$, a harmonic floor that remains active even when the spatial momentum approaches zero. In the nonrelativistic regime, where the kinetic energy is heavily subordinated to the rest mass scale, a clean separation of scales can be achieved by factoring out this fast-oscillating fundamental carrier.

To formalize this approximation, we decouple the rapid phase carrier from the slower dynamical modulations by expressing the full field as:
\begin{equation}
	\psi(x,t) = e^{-i\omega_0 t}\,\phi(x,t),
	\label{eq:phase_factorization}
\end{equation}
where the envelope function $\phi(x,t)$ is assumed to vary slowly in time relative to the rapid temporal scale dictated by $\omega_0$.

\subsection{Expansion of the Klein--Gordon equation}

To trace the dynamical consequences of this separation, we insert the factored ansatz~(\ref{eq:phase_factorization}) into the field equation~(\ref{eq:kg_nr_start}). Evaluating the temporal and spatial differential actions on the field yields the expanded expressions:
\begin{align}
	\frac{\partial \psi}{\partial t} &= e^{-i\omega_0 t}\left(-i\omega_0 \phi + \frac{\partial \phi}{\partial t}\right), \\
	\frac{\partial^2 \psi}{\partial t^2} &= e^{-i\omega_0 t}\left(-\omega_0^2 \phi - 2i\omega_0 \frac{\partial \phi}{\partial t} + \frac{\partial^2 \phi}{\partial t^2}\right),
\end{align}
along with the corresponding spatial Laplacian:
\begin{equation}
	\nabla^2 \psi = e^{-i\omega_0 t}\nabla^2 \phi.
\end{equation}

Reconstructing these terms within the definition of the d'Alembert operator, $\Box \psi \equiv \frac{1}{c^2}\frac{\partial^2 \psi}{\partial t^2} - \nabla^2 \psi$, allows the full invariant differential action to be written as:
\begin{align}
	\Box \psi = e^{-i\omega_0 t} \Bigg[
	&-\frac{\omega_0^2}{c^2}\phi
	- \frac{2i\omega_0}{c^2}\frac{\partial \phi}{\partial t}
	+ \frac{1}{c^2}\frac{\partial^2 \phi}{\partial t^2}
	- \nabla^2 \phi
	\Bigg].
\end{align}

Substituting this structured differential expansion into the core Klein--Gordon equation~(\ref{eq:kg_nr_start}) yields:
\begin{align}
	e^{-i\omega_0 t} \Bigg[
	&-\frac{\omega_0^2}{c^2}\phi
	- \frac{2i\omega_0}{c^2}\frac{\partial \phi}{\partial t}
	+ \frac{1}{c^2}\frac{\partial^2 \phi}{\partial t^2}
	- \nabla^2 \phi
	+ \frac{\omega_0^2}{c^2}\phi
	\Bigg] = 0.
\end{align}
Crucially, direct algebraic cancellation eliminates the static frequency terms $-\frac{\omega_0^2}{c^2}\phi$ and $+\frac{\omega_0^2}{c^2}\phi$, leaving the intermediate envelope equation:
\begin{equation}
	-\frac{2i\omega_0}{c^2}\frac{\partial \phi}{\partial t}
	+ \frac{1}{c^2}\frac{\partial^2 \phi}{\partial t^2}
	- \nabla^2 \phi = 0.
	\label{eq:intermediate_phi}
\end{equation}

\subsection{Nonrelativistic approximation}

The nonrelativistic limit is formally defined by the requirement that the envelope $\phi$ evolves on a time scale much longer than the underlying clock period $\omega_0^{-1}$. This physical slow-variation condition implies the inequality:
\begin{equation}
	\left|\frac{\partial^2 \phi}{\partial t^2}\right| \ll \omega_0 \left|\frac{\partial \phi}{\partial t}\right|.
\end{equation}
Consequently, the second-order temporal derivative in Eq.~(\ref{eq:intermediate_phi}) can be safely neglected, simplifying the dynamical constraint to:
\begin{equation}
	-\frac{2i\omega_0}{c^2}\frac{\partial \phi}{\partial t}
	- \nabla^2 \phi = 0.
\end{equation}

Isolating the first-order time derivative straightforwardly yields:
\begin{equation}
	i \frac{\partial \phi}{\partial t}
	= -\frac{c^2}{2\omega_0} \nabla^2 \phi.
\end{equation}

Finally, substituting the structural mass--frequency identity $\hbar \omega_0 = mc^2$ transforms this relation into:
\begin{equation}
	i\hbar \frac{\partial \phi}{\partial t}
	= -\frac{\hbar^2}{2m} \nabla^2 \phi,
	\label{eq:schrodinger}
\end{equation}
which matches exactly the standard Schrödinger equation for a free nonrelativistic particle \cite{greiner1990}.

\subsection{Physical interpretation}

The seamless emergence of the Schrödinger equation from the underlying Klein--Gordon structure carries a transparent physical interpretation within this framework. The complete field $\psi$ represents a fundamentally relativistic entity whose rapid phase factor $e^{-i\omega_0 t}$ tracks the intrinsic proper frequency of the excitation. The wave function $\phi(x,t)$, on the other hand, merely describes the slowly varying envelope that modulates this background clock.

Viewed from this perspective, nonrelativistic quantum mechanics reveals itself as a specific dynamical regime: one where the dominant contribution to phase evolution is dictated by the field's intrinsic rest periodicity, while standard nonrelativistic quantum phenomena emerge as small spatial and temporal deviations tracked by the modulation envelope $\phi$.

As a result, the Schrödinger equation does not require an independent, nonrelativistic axiomatic postulate. It arises naturally as the effective low-energy equation governing the slow envelope dynamics of a relativistic field endowed with intrinsic phase periodicity. The mass parameter appearing in Eq.~(\ref{eq:schrodinger}) is inherited directly from the foundational rest frequency scale $\omega_0$, reinforcing the structural view of mass as an operational measure of localized field oscillation and completing the bridge between relativistic field dynamics and nonrelativistic wave mechanics.

\section{Propagation in External Potentials}

\subsection{Step and barrier potentials}

To examine how external constraints modify this phase architecture, let us introduce a static scalar potential $V(x)$. At a phenomenological level, interactions are conventionally incorporated via the standard gauge-like minimal prescription:
\begin{equation}
	i\hbar \frac{\partial}{\partial t} \rightarrow i\hbar \frac{\partial}{\partial t} - V(x),
\end{equation}
which generalizes the free field dynamics into the wave equation:
\begin{equation}
	\left[
	\frac{1}{c^2}\left(i\hbar \frac{\partial}{\partial t} - V(x)\right)^2
	- \hbar^2 \nabla^2
	- m^2 c^2
	\right]\psi = 0.
	\label{eq:kg_potential}
\end{equation}
While this specific prescription is not manifestly covariant due to the static nature of the background field, it provides a highly reliable effective description for stationary external configurations and allows a direct, transparent bridge to the nonrelativistic limit \cite{greiner1990}.

Restricting our analysis to stationary state configurations of the form,
\begin{equation}
	\psi(x,t) = \phi(x)e^{-iEt/\hbar},
\end{equation}
where $E$ denotes the total energy, direct substitution into Eq.~(\ref{eq:kg_potential}) yields the spatial differential equation:
\begin{equation}
	\left[
	\frac{(E - V(x))^2}{c^2}
	+ \hbar^2 \frac{d^2}{dx^2}
	- m^2 c^2
	\right]\phi(x) = 0.
\end{equation}
This can be written compactly as a standard Helmholtz-like equation:
\begin{equation}
	\frac{d^2 \phi}{dx^2} + k^2(x)\phi = 0,
	\label{eq:spatial_eq}
\end{equation}
where the effective local wavevector profile is explicitly defined by:
\begin{equation}
	k^2(x) = \frac{(E - V(x))^2 - m^2 c^4}{\hbar^2 c^2}.
	\label{eq:k_definition}
\end{equation}

As a primary testing ground for this interacting phase structure, consider a sharp step potential profile:
\begin{equation}
	V(x) =
	\begin{cases}
		0, & x < 0, \\
		V_0, & x \ge 0.
	\end{cases}
\end{equation}
Within each piecewise region, the field dynamics simplifies to a constant-coefficient differential equation, whose solutions branch into either purely oscillatory plane waves or exponential envelopes depending entirely on the algebraic sign of $k^2(x)$.

\subsection{Evanescent waves and tunneling}
\label{sec:tunneling}
The spatial mapping of the field is fundamentally governed by the local sign of the squared wavevector $k^2(x)$, splitting the space into distinct propagation regimes.

Evaluating the system within the upstream region ($x < 0$), where the potential vanishes ($V=0$), the squared wavevector reads:
\begin{equation}
	k_1^2 = \frac{E^2 - m^2 c^4}{\hbar^2 c^2}.
\end{equation}
For an incoming state above the rest mass threshold ($E > mc^2$), the parameter $k_1^2$ remains strictly positive, yielding a standard oscillatory superposition of traveling waves:
\begin{equation}
	\phi_I(x) = A e^{i|k_1| x} + B e^{-i|k_1| x}.
\end{equation}

Conversely, within the downstream step region ($x \ge 0$), the potential shifts the local wavevector expression to:
\begin{equation}
	k_2^2 = \frac{(E - V_0)^2 - m^2 c^4}{\hbar^2 c^2}.
\end{equation}
Whenever the total energy falls within the bounded energy window,
\begin{equation}
	|E - V_0| < mc^2,
\end{equation}
the squared wavevector turns negative ($k_2^2 < 0$). Defining the real attenuation parameter,
\begin{equation}
	\kappa = \sqrt{\frac{m^2 c^4 - (E - V_0)^2}{\hbar^2 c^2}},
\end{equation}
the spatial modulation profile shifts from traveling to evanescent, requiring a physically regularized exponential decay:
\begin{equation}
	\phi_{II}(x) = C e^{-\kappa x}.
\end{equation}
This confirms that inside the classically forbidden domain, the material field preserves a non-vanishing, exponentially bounded amplitude.

This evanescent architecture underpins the wave-mechanical description of relativistic tunneling. For a finite potential barrier of spatial width $L$, defined by the profile:
\begin{equation}
	V(x) =
	\begin{cases}
		0, & x < 0, \\
		V_0, & 0 \le x \le L, \\
		0, & x > L,
	\end{cases}
\end{equation}
the solution within the barrier domain ($0 \le x \le L$) retains both exponential degrees of freedom:
\begin{equation}
	\phi_{II}(x) = C e^{-\kappa x} + D e^{\kappa x}.
\end{equation}
Enforcing standard continuity matching conditions for the field and its first spatial derivative across the boundaries at $x=0$ and $x=L$ straightforwardly yields a non-vanishing transmission coefficient. Even when the wavevector becomes imaginary within the barrier region, the field maintains phase and amplitude coherence across the obstacle, enabling partial transmission into the downstream free region.

\subsection{Interpretation as propagation below cutoff}

The physical condition that sustains oscillatory propagation within an interacting region can be rewritten by mapping energy and mass parameters onto their corresponding wave frequencies. Setting $E = \hbar \omega$ and $mc^2 = \hbar \omega_0$, the traveling wave requirement $(E - V)^2 > m^2 c^4$ assumes the transparent frequency form:
\begin{equation}
	(\omega - V/\hbar)^2 > \omega_0^2.
\end{equation}

This representation reveals that an external potential acts operationally by locally shifting the field's frequency spectrum. Traveling, propagating wave modes can only be sustained when this effective, shifted operating frequency strictly exceeds the field's intrinsic rest frequency $\omega_0$. Conversely, in regions where the interaction forces the system below this threshold,
\begin{equation}
	(\omega - V/\hbar)^2 < \omega_0^2,
\end{equation}
the wavevector turns imaginary, forcing the field to decay exponentially as an evanescent mode.

This behavior maps identically onto the classical physics of wave propagation inside dispersive media or wave-guiding structures endowed with a sharp cutoff threshold. The invariant proper frequency $\omega_0$ represents a fundamental, built-in cutoff scale below which traveling modes are strictly forbidden. The external potential dynamically alters the local frequency landscape, shifting the field's operating frequency relative to this baseline threshold. 

Viewed through this lens, quantum tunneling loses its paradoxical character. Rather than invoking point particles that temporarily violate energy conservation laws to bridge a barrier, tunneling emerges as the natural transmission of an evanescent wave through a medium whose local cutoff frequency temporarily exceeds the operating frequency of the incoming field. This formulation unifies relativistic quantum tunneling and classical wave mechanics under a single, cohesive principle: mass acts as an intrinsic frequency scale, and interactions merely distort the underlying phase-propagation landscape.

\section{Discussion}
\label{sec:discussion}

\subsection{Mass as intrinsic frequency}

The theoretical formulation developed in this work offers a direct, operational interpretation of mass strictly in terms of intrinsic phase periodicity. The fundamental relation,
\begin{equation}
	\hbar \omega_0 = mc^2
\end{equation}
maps the mass scale of a material excitation onto a proper-time oscillation frequency, preventing it from being introduced as an isolated, non-wave dynamical parameter.

Under this conceptual inversion, rest energy ceases to be an abstract kinematic property; it represents the energy content structurally bound to an internal periodic phenomenon. The presence of a non-vanishing mass guarantees that even when spatial momentum is entirely removed, the field preserves a non-trivial temporal architecture. Mass, in this sense, acts as the foundational metric that parameterizes the intrinsic time scale of the excitation.

This perspective remains fully consistent with the phase-centric kinematics established in the preceding sections, where proper time operationally tracks the accumulation of phase cycles along a system's worldline. The mass parameter emerges naturally as the structural coupling constant that anchors phase evolution to proper time, providing a unified conceptual baseline that merges relativistic kinematics and local field dynamics into a single reality.

\subsection{Wave interpretation of relativistic quantum dynamics}

The Klein--Gordon equation, reconstructed here from foundational phase periodicity constraints, yields a highly intuitive parallel with the physics of wave propagation through a dispersive medium endowed with a sharp cutoff threshold $\omega_0$. Within this wave-mechanical framework, relativistic quantum dynamics is fully described by the collective behavior of wave packets governed by the dispersion relation:
\begin{equation}
	\omega^2 = c^2 k^2 + \omega_0^2.
\end{equation}

The group velocity of these wave envelopes maps identically onto the relativistic velocity of a classical particle, while the lower bound enforced by a non-zero $\omega_0$ dictates natural dispersion and prevents the stability of infinitely localized states. As demonstrated in Sec.~\ref{sec:tunneling}, external potentials distort this effective frequency landscape, setting up sharp boundaries between traveling and evanescent regimes, which strips quantum tunneling of its non-intuitive or paradoxical character.

Furthermore, in the nonrelativistic limit, isolating and removing the rapid proper-time carrier frequency straightforwardly yields the Schrödinger equation for the slowly varying envelope. In this low-energy regime, nonrelativistic quantum mechanics reveals its true structural origin: it behaves as an effective modulation framework tracking minor perturbations over a deeply relativistic, highly oscillating baseline phase. Taken together, these results demonstrate that particle-like behaviors are emergent phenomena, fully subordinated to the interference properties of fields endowed with intrinsic periodicity.

\subsection{Relation to standard formulations}

It is worth noting that this phase-based formulation leaves the empirical predictions of relativistic quantum mechanics completely unaltered. Instead, it reorders the underlying logical hierarchy. The Klein--Gordon equation is no longer presented as an ad-hoc operator substitution applied to a classical energy-momentum relation; it emerges as the minimal local, linear, and covariant differential equation capable of sustaining intrinsic phase coherence.

This approach represents the natural field-theoretic extension of a recently developed framework wherein special relativity is reconstructed from phase invariance across moving frames \cite{puddu2026kinematics}. While that foundational work demonstrated that relativistic energy and momentum formulas arise from phase synchronization conditions, the present paper confirms that the exact same principles dictate the structural form of the field equations. In textbook formulations, mass enters the Lagrangian or the field equations as an arbitrary phenomenological coupling constant \cite{peskin1995}. Here, the same algebraic term is recognized as the geometric footprint of an invariant frequency, preserving the established mathematical apparatus while providing a more transparent physical rationale.

A cautionary remark is in order regarding the introduction of interactions in Sec.~\ref{sec:tunneling}, which was executed at a phenomenological level through a static scalar potential. A more rigorous, first-principles derivation of interactions within a phase-coherence framework remains an open question and will likely require additional geometric postulates. Similarly, while identifying mass with an internal frequency scale unifies several conceptual domains, it does not by itself explain the physical origin of this fundamental clock rate. This work should therefore be viewed as a robust structural reinterpretation of relativistic wave equations rather than a complete, closed dynamical theory.

Finally, to address these limitations and move beyond a purely scalar description, the framework must be compatible with fully gauge-covariant structures. If extended to charged fields, external electromagnetic interactions cannot be introduced solely through a non-covariant scalar potential; they must incorporate the standard four-potential $A_\mu=(\phi/c,-\mathbf{A})$ via the minimal covariant derivative substitution \cite{peskin1995}:
\begin{equation}
	\partial_\mu \rightarrow D_\mu=\partial_\mu+\frac{iq}{\hbar}A_\mu .
\end{equation}
In this generalized configuration, the field's underlying phase geometry is coupled directly to the electromagnetic gauge connection, rendering the total accumulated phase path-dependent in a standard gauge-theoretic sense. Developing the full mathematical machinery for this spinning and charged extension represents the necessary next step for future investigations.

\section{Conclusions}
\label{sec:conclusions}

This work has developed a coherent, phase-centric formulation of relativistic wave dynamics, demonstrating that the Klein--Gordon equation can be systematically derived by demanding intrinsic phase periodicity from material fields. By mapping the wave phase directly onto the classical action and introducing an invariant rest-frame frequency $\omega_0$, we have shown that the mass scale emerges as a structural identity through the relation $\hbar \omega_0 = mc^2$. This approach eliminates the traditional need to inject rest energy into the theory as an independent, axiomatic starting input.

Under this logical hierarchy, the Klein--Gordon structure ceases to be an ad-hoc quantum postulate; it manifests as the minimal local, linear, and Lorentz-invariant field architecture capable of sustaining this foundational periodic constraint. The resulting field solutions govern the propagation of waves characterized by a non-vanishing proper-time oscillation floor, straightforwardly determining a relativistic dispersion relation bounded by a sharp cutoff frequency.

We have explored the physical and dynamical implications of this formulation across multiple operational regimes. In free propagation, the standard relativistic mapping between group velocity and classical particle kinematics is recovered as an emergent property of wave interference, while localized wave packet envelopes undergo natural spatial dispersion controlled entirely by the intrinsic frequency scale. In the low-energy regime, a systematic separation of scales that filters out the rapid proper-time carrier successfully recovers the Schrödinger equation for the slowly varying envelope, framing nonrelativistic mechanics as an effective modulation theory. Furthermore, when evaluating interactions with external potentials, the sharp transition between traveling and evanescent states strips relativistic tunneling of its non-intuitive character, reinterpreting it as simple wave propagation below a local cutoff threshold.

Collectively, these insights suggest that the algebraic and kinematic architecture of relativistic quantum dynamics can be fully understood through the lens of wave propagation governed by intrinsic periodicity. Within this framework, particle-like attributes are recognized as collective, secondary properties arising from the dispersive structure of the underlying fields. While the present analysis has been restricted to scalar fields and phenomenological potentials, it establishes a mathematically rigorous and conceptually unified baseline that bridges wave phase, mass thresholds, and relativistic dynamics.

Natural avenues for future inquiry include the first-principles derivation of interactions at a more foundational, gauge-covariant level, alongside the extension of these phase-coherence principles to fields endowed with internal degrees of freedom, such as spin. More broadly, elevating intrinsic phase periodicity to a fundamental organizing principle may provide a useful perspective for exploring the structural connections that bind wave mechanics to the underlying geometry of spacetime.
	
	%\begin{acknowledgments}
	%\end{acknowledgments}
	
	\bibliographystyle{apsrev4-2}
	\bibliography{references}

@article{debroglie1924,
  author = {de Broglie, Louis},
  title = {Recherches sur la théorie des quanta},
  journal = {Philosophical Magazine},
  volume = {47},
  pages = {446--458},
  year = {1924}
}

@article{klein1926,
  author = {Klein, Oskar},
  title = {Quantentheorie und fünfdimensionale Relativitätstheorie},
  journal = {Zeitschrift für Physik},
  volume = {37},
  pages = {895--906},
  year = {1926}
}

@article{gordon1926,
  author = {Gordon, Walter},
  title = {Der Comptoneffekt nach der Schrödingerschen Theorie},
  journal = {Zeitschrift für Physik},
  volume = {40},
  pages = {117--133},
  year = {1926}
}

@book{peskin1995,
  author = {Peskin, M. E. and Schroeder, D. V.},
  title = {An Introduction to Quantum Field Theory},
  publisher = {Westview Press},
  year = {1995}
}

@book{greiner1990,
  author = {Greiner, W.},
  title = {Relativistic Quantum Mechanics: Wave Equations},
  publisher = {Springer},
  year = {1990}
}

@article{muller2014,
  author = {Müller, Holger},
  title = {Quantum mechanics, matter waves and moving clocks},
  journal = {Nature Physics},
  volume = {10},
  pages = {727--731},
  year = {2014}
}

@article{kragh1984,
  author = {Kragh, Helge},
  title = {Equation with the many fathers: The Klein--Gordon equation in 1926},
  journal = {American Journal of Physics},
  volume = {52},
  pages = {1024--1033},
  year = {1984}
}

@misc{puddu2026kinematics,
  author = {Emiliano Puddu},
  title = {Relativistic Kinematics from Phase Coherence and Intrinsic Periodicity},
  note = {Submitted to Foundations of Physics, under review},
  year = {2026}
}
	% Suggested references (fill publisher/year/edition as desired)

\end{document}